\documentclass[reprint,amsmath,amssymb,aps,prb]{revtex4-2}

\usepackage{adjustbox}
\usepackage{placeins}
\usepackage{newtxtext}
\usepackage[varvw]{newtxmath}
\usepackage{graphicx}
\usepackage{dcolumn}

\usepackage{bm}
\usepackage{bbm}
\usepackage{csquotes}
\usepackage{mathtools}
\usepackage{booktabs,makecell}
\usepackage{physics}
\usepackage{subfigure}
\usepackage{tikz}

\usepackage{amsfonts,amsmath,amssymb,amsthm}
\usepackage[colorlinks=true,allcolors=blue,breaklinks=True]{hyperref}

\newcommand{\klr}[1]{\left(#1\right)}
\newcommand{\kle}[1]{\left[#1\right]}

\newcommand{\klsr}[1]{#1 \biggl|}
\newcommand{\kla}[1]{\langle#1\rangle}
\newcommand{\klg}[1]{\left\{#1\right\}}

\newcommand{\varvec}[1]{\mathbf{#1}}
\renewcommand{\vv}[1]{\varvec{#1}}

\newcommand{\svec}[1]{\begin{pmatrix}#1\end{pmatrix}}
\newcommand{\matr}[1]{\svec{#1}}
\newcommand{\varvarvec}[1]{\bm{#1}}
\newcommand{\vvv}[1]{\varvarvec{#1}}

\newcommand{\fermifunc}{n_{\mathrm{F}}}



\begin{document}

\title{Low-temperature scaling laws in unconventional flat-band superconductors}

\author{Maximilian Buthenhoff}
\email{buthenhoff.m.b890@m.isct.ac.jp}
\affiliation{Department of Physics, Institute of Science Tokyo, Ookayama, Meguro, Tokyo 152-8551, Japan}

\author{Yusuke Nishida}
\affiliation{Department of Physics, Institute of Science Tokyo, Ookayama, Meguro, Tokyo 152-8551, Japan}

\date{\today}

\begin{abstract}
In flat-band superconductors, the electron pairing is strongly enhanced so that the critical temperature scales linearly with the interaction strength. Identifying the governing pairing mechanism in flat-band superconducting systems is therefore a central task, which may be constrained by experimental probes via low-temperature scaling measurements. A key observable underlying the Meissner effect and the resulting divergent dc conductivity is the superfluid weight. While it is well established that the minimal quantum metric provides the dominant contribution to the superfluid weight in conventional superconductors with isolated flat bands, recent studies indicate that the unconventional pairing can generate additional nonlocal quantum geometric terms. This motivates us to derive the low-temperature scaling law of the superfluid weight in two-dimensional flat-band superconductors with sufficiently isolated bands. In particular, we consider the gap function with point or line nodes classified by the Weierstrass preparation theorem. Beyond the superfluid weight, we additionally deliver explicit low-temperature scaling laws of the order parameter, the tunneling conductance, the specific heat, the Sommerfeld coefficient, and the spin-lattice relaxation rate to provide complementary experimental discriminants of the underlying pairing symmetry. The implications of our results are also elucidated by applying them to a selection of superconducting states in $C_{6v}$-symmetric systems.
\end{abstract}

\maketitle

\section{Introduction}
Because of the divergent density of states in a flat band, flat-band superconductors represent promising candidates for high-temperature superconductors~\cite{kopnin2011high,heikkila2016flat,balents2020superconductivity,andrei2021marvels,bouzerar2022giant,layek2023possible,nunez2024magnetic,thumin2025crossing}. To identify the possible pairing mechanism, it is beneficial to have a good understanding of the underlying nodal structure~\cite{annett1990symmetry,sigrist1991phenomenological}. One experimentally accessible method for the identification of the correct nodal structure is the measurement of the low-temperature scaling of observables~\cite{annett1991interpretation,wu1993temperature,mazidian2013anomalous,khvalyuk2024near,tanaka2024superfluid}. \\ 
\indent When ignoring interband pairing (which gives rise to offsets in the scaling laws of the density of states determined by the strength of a pseudo-magnetic field~\cite{lapp2020experimental}), the low-temperature scaling laws of observables are completely fixed by the topology of the zeros, i.e., nodes, of the gap function in momentum space~\cite{volovik1989.49.685,sigrist1991phenomenological}. One key quantity that indicates the possibility of superconductivity is the superfluid weight (or superfluid stiffness) \cite{london1935electromagnetic}. In the absence of Galilean invariance, it has been shown that nontrivial single-particle quantum geometry characterizes the underlying mechanism responsible for the existence of a nonzero superfluid weight in a flat band with divergent effective mass \cite{khodel1990superfluidity,zeng2025superfluid}. While the minimal quantum metric is solely responsible for a nonzero superfluid weight in conventional superconductors with isolated flat bands \cite{huhtinen2022revisiting}, recent studies indicate the appearance of nonlocal quantum-geometric terms in the superfluid weight for the majority of unconventional pairing due to the momentum-dependent nature of the gap function~\cite{zeng2025superfluid,lamponen2025superconductivity,buthenhoff2025functional}. Therefore, it is important to clarify whether the scaling laws obtained in Ref.~\cite{hirobe2025anomalous} for the superfluid weight of flat-band superconductors need to be adjusted when taking into account the arising functional, i.e., nonlocal quantum geometrical, superfluid weight. \\
\indent For completeness, we additionally calculate the low-temperature scaling laws of the order parameter, the tunneling conductance, the specific heat, the Sommerfeld coefficient, and the spin-lattice relaxation rate for flat-band superconductors with nontrivial nodal structure, providing a full guide for experimental measurements. Our main results are summarized in Table \ref{tab:main-result}. See also Ref.~\cite{putzer2025eliashberg}, where the authors discuss the low-temperature behavior of the superfluid weight in twisted graphene in the presence of band-off-diagonal pairing, and Ref.~\cite{penttila2025flat}, in which dynamical mean-field theory calculations for an attractive Hubbard model on the Lieb lattice indicate that the superfluid weight follows a Gorter-Casimir-like behavior~\cite{tinkham2004introduction,pond1991penetration}. 

\begin{table*}[t!]
\caption{Collection of low-temperature scaling laws in flat-band superconductors with different nodal structures covered by the dispersion~\eqref{eq:general-dispersion} for the order parameter $\Delta_T$, the geometrical and functional superfluid weights $D_{\mathrm{s}}^{\mathrm{geom},\mathrm{func}}$, the tunneling conductance $G_{sn}$, the specific heat $C$, the Sommerfeld coefficient $\gamma$, and the NMR spin-lattice relaxation rate $1/(T_1T)$. Here, $m>0$ represents the total order of vanishing of the gap function and $L$ indicates the number of straight nodal lines through the origin, i.e., the case of $L = 1$ describes a single line node without crossings, $L = 2$ a crossing of two line nodes, and $L > 2$ crossings of three or more line nodes. Moreover, $b \ge 0$ represents a superconducting state dependent exponent which is determined by Eq.~\eqref{eq:gamma-exponent-definition}. We have $b=1$ for all one-dimensional superconducting states as well as for all two-dimensional states considered in Sec.~\ref{sec:application}.}\label{tab:main-result}
\renewcommand{\arraystretch}{1.4}
\begin{ruledtabular}
\begin{tabular}{llllllll}
 node type & $\Delta_T$ & $D_{\mathrm{s}}^{\mathrm{geom}}$ & $D_{\mathrm{s}}^{\mathrm{func}}$ & $G_{sn}$ & $C$ & $\gamma$ & $1/(T_1T)$ \\ 
\colrule
    point node & $T^{\frac{2}{m} +1}$ & $T^{\frac{2}{m} +1}$ & $T^{{\frac{2}{m} + b}}$ & $T^{\frac{2}{m} - 1}$ & $T^{\frac{2}{m}}$ & $T^{\frac{2}{m} - 1}$ & $T^{\frac{4}{m} - 2}$ \\
    line node ($L = 1$) & $T^{\frac{1}{m} + 1}$ & $T^{\frac{1}{m} + 1}$ & $T^{{\frac{1}{m} + b}}$ &  $T^{\frac{1}{m} - 1}$ & $T^{\frac{1}{m}}$ & $T^{\frac{1}{m} - 1}$ & $T^{\frac{2}{m} - 2}$ \\
    line node ($L = 2$) & $T^{\frac{2}{m} +1} \ln(1/T)$ & $T^{\frac{2}{m} +1}\ln(1/T)$ & $T^{{\frac{2}{m} + b}}\ln(1/T)$ & $T^{\frac{2}{m} - 1}\ln(1/T)$ & $T^{\frac{2}{m}}\ln(1/T)$ & $T^{\frac{2}{m} - 1}\ln(1/T)$ & $T^{\frac{4}{m} - 2}\ln^2(1/T)$ \\
    line node ($L > 2$) & $T^{\frac{2}{m} +1}$ & $T^{\frac{2}{m} +1}$ & $T^{{\frac{2}{m} + b}}$ & $T^{\frac{2}{m} - 1}$ & $T^{\frac{2}{m}}$ & $T^{\frac{2}{m} - 1}$ & $T^{\frac{4}{m} - 2}$ 
\end{tabular}
\end{ruledtabular}
\end{table*}

This paper is organized as follows. In Sec.~\ref{sec:nodal-structure}, we discuss possible nodal structures of the gap function in two-dimensional flat-band superconductors by utilizing the Weierstrass preparation theorem~\cite{Weierstrass1895,Golubitsky1973,Krantz2002}. In particular, Eq.~\eqref{eq:general-dispersion} presents the dispersion we work with throughout this work. In Sec.~\ref{sec:dos}, we derive the density of states for each case covered by the dispersion. Sec.~\ref{sec:scaling-laws} collects derivations of the low-temperature scaling laws of the order parameter, the geometrical and functional superfluid weights, the tunneling conductance, the specific heat, the Sommerfeld coefficient, and the NMR spin-lattice relaxation rate. Lastly, we apply our results to the case of $C_{6v}$-symmetric systems in Sec.~\ref{sec:application}.
\section{Nodal structure of the gap function}\label{sec:nodal-structure}
The grand potential of mean-field BCS theory for nonzero temperature is given by \cite{huhtinen2022revisiting} 
\begin{align}
    \Omega(\vv{q}) &= -T \sum_{\vv{k},n} \ln\left[1 + \exp\klr{-\frac{E_{\vv{k}n}(\vv{q})}{T}}\right] + \sum_{\vv{k}} \tr(\varepsilon_{\vv{k}-\vv{q}} - \mu\mathbbm{1}) \nonumber \\
    &+ \frac{V}{2} \sum_{\vv{k},\vv{k}'} U^{-1}(\vv{k},\vv{k}') \Delta^\dagger_{\alpha\beta}(\vv{q};\vv{k}) \Delta_{\beta\alpha}(\vv{q};\vv{k}') \,,
    \label{eq:grand-potential-mf-bcs}
\end{align}
where we set $k_{\mathrm{B}} = e = 1.$ Here, $E_{\vv{k}n}(\vv{q})$ represents the eigenvalue with index $n = 1,\hdots,2N_{\mathrm{B}}$ of the Bogolioubov-de Gennes (BdG) Hamiltonian at momentum $\vv{k}$ in the presence of an external gauge field $\vv{q} = \vv{A}$,
\begin{align}
    \mathcal{H}_{\mathrm{BdG}}(\vv{k},\vv{q}) = \matr{H(\vv{k} - \vv{q}) - \mu\mathbbm{1} & \Delta(\vv{q};\vv{k})\\ \Delta^\dagger(\vv{q};\vv{k}) & -H(\vv{k} + \vv{q}) + \mu\mathbbm{1}} \,,
\end{align}
$T$ represents the temperature, $\varepsilon_{\vv{k}} = \mathrm{diag}(\varepsilon_{\vv{k}1},\hdots,\varepsilon_{\vv{k}N_{\mathrm{B}}})$ contains the $N_{\mathrm{B}}$ eigenvalues of the time-reversal symmetric~(TRS) single-particle Hamiltonian $H(\vv{k})$, $\mu$ is the chemical potential, $U(\vv{k},\vv{k}')$ is the complex-valued effective pairing potential, and $\Delta_{\alpha\beta}(\vv{q};\vv{k})$ is the gap function with band indices $\alpha,\beta = 1,\hdots,N_{\mathrm{B}}$ which contains the order parameter. Note that the formalism of this work allows the consideration of TRS-breaking pairing mechanisms, but not TRS-breaking single-particle Hamiltonians. If the single-particle Hamiltonian $H(\vv{k})$ exhibits a flat band, its time-reversed partner has the same spectrum (up to $\vv{k} \to -\vv{k}$), and hence, exhibits the same flat band. Consequently, the DOS and the corresponding scaling laws of the order parameter, the tunneling conductance, the specific heat, the Sommerfeld coefficient, and the NMR relaxation rate in Table~\ref{tab:main-result} are unchanged. However, the time-reversed single-particle eigenstates do change. This leads to different and additional quantum geometric quantities within the superfluid weight \cite{huhtinen2022revisiting}, potentially with a different low-temperature scaling, which can modify Table~\ref{tab:main-result} in the TRS-breaking case. \\
\indent The classification of superconducting states is done via the irreducible representations of the symmetry group of the system \cite{mineev1999introduction}. In particular, we need to distinguish between spin-singlets with total spin $S = 0$ and spin-triplets with total spin $S = 1$. The pair spin wave function of a spin-singlet is antisymmetric with respect to an exchange of the spin indices, i.e., the gap function can be expressed as
\begin{align}
    \Delta_{\alpha\beta}(\vv{k}) = f_{\alpha\beta}(\vv{k})i\sigma_y \,, \quad f_{\alpha\beta}(\vv{k}) = f_{\beta\alpha}(-\vv{k}) \,,
\end{align}
where $\Delta_{\alpha\beta}(\vv{k}) = \Delta_{\alpha\beta}(\vv{q}=0;\vv{k})$ and the function $f_{\alpha\beta}$ is determined by the symmetry group of the system. If $\Gamma$ denotes an irreducible representation of the group with dimension $d_\Gamma$, it is provided by
\begin{align}
    f_{\alpha\beta}(\vv{k}) = \sum_{i=1}^{d_\Gamma} \Delta_{\alpha\beta}^i \psi_i^{\Gamma}(\vv{k}) \,,
    \label{eq:form-factor-definition}
\end{align}
where $\psi_i^{\Gamma}$ are the simplest basis functions that are even in $\vv{k}$ and respect the symmetry of the system and $\Delta_{\alpha\beta}^i$ are coefficients representing the order parameters of the superconductor. Similarly, a spin-triplet state with total spin $S=1$ has odd parity, i.e., the gap function can be expressed as  
\begin{align}
    \Delta_{\alpha\beta}(\vv{k}) = (\vv{d}_{\alpha\beta}(\vv{k}) \cdot \vvv{\sigma})i\sigma_y\,, \quad 
    \vv{d}_{\alpha\beta}(\vv{k}) = -\vv{d}_{\beta\alpha}(-\vv{k})\,.
\end{align}
In absence of spin-orbit coupling, the function $\vv{d}_{\alpha\beta}$ is given by
\begin{align}
    \vv{d}_{\alpha\beta}(\vv{k}) = \hat{\vv{n}} \sum_{i=1}^{d_\Gamma} \Delta^i_{\alpha\beta} \psi_i^\Gamma(\vv{k}) \equiv \hat{\vv{n}} f_{\alpha\beta}(\vv{k}) \,,
\end{align}
where $\hat{\vv{n}}$ is a fixed spin direction \cite{mineev1999introduction}. Note that, unlike in the spin-singlet case, $f_{\alpha\beta}$ is of odd parity for spin-triplets. \\
\indent In the following, we assume that the pairing potential associated to $\Gamma$ factorizes such that it can be expressed as \cite{mineev1999introduction}
\begin{align}
    U(\vv{k},\vv{k}') = U_0 \sum_{i=1}^{d_\Gamma} \psi_i^\Gamma(\vv{k}') \bar{\psi}_i^\Gamma(\vv{k}) \,.
    \label{eq:pairing-potential}
\end{align}
Under this assumption, the self-consistent equations for the order parameters can be written as 
\begin{align}
    \Delta_{\alpha\beta}^i = \frac{U_0}{V} \sum_{\vv{k},n} \bar{\psi}_i^\Gamma(\vv{k}) v_{n\alpha,+}(\vv{k}) v^\ast_{n\beta,-}(\vv{k}) [1 - 2\fermifunc(E_{\vv{k}n})]
    \label{eq:self-consistent-eq-general}
\end{align}
at $\vv{q} = 0$. Here, the $2N_{\mathrm{B}}$ vectors $(v_{n,+},v_{n,-})$ represent the eigenvectors of the BdG Hamiltonian in the Nambu spinor basis. Moreover, we separate the overall temperature dependence of the gap function from its internal structure within the $d_\Gamma$-dimensional space of the irreducible representation. Concretely, we assume that all nonvanishing components share a single temperature-dependent amplitude $\Delta_T$ and differ only by fixed temperature-independent coefficients $\eta_i$, i.e., $\Delta^{i}_{\alpha\beta} \equiv \Delta_T \eta_i \delta_{\alpha\beta}$ for each $i = 1,\hdots,d_{\Gamma}$, and we restrict $(\eta_1,\hdots,\eta_{d_{\Gamma}})$ to a set of high-symmetry representative directions where each component is either absent or present with unit modulus. Examples include nematic, chiral, or cyclic states \cite{ohashi2024surface}. For a given choice of $(\eta_1,\hdots,\eta_{d_{\Gamma}})$, the amplitude $\Delta_T$ and its temperature dependence is determined by the self-consistent equation,
\begin{align}
    \Delta_T{\eta}_i = \frac{U_0}{2V} \sum_{\vv{k},n} \frac{\bar{\psi}_i^\Gamma(\vv{k}) f(\vv{k})}{E_{\vv{k}n}} [1 - 2n_{\mathrm{F}}(E_{\vv{k}n})] \,,
    \label{eq:self-consistent-eq}
\end{align}
for a fixed $i$ with $\eta_i \neq 0$, while the remaining equations provide a consistency check. In the following, we absorb the phase factors $\eta_i \neq 0$ into the basis functions. \\
\indent The quasiparticle dispersion in a flat band can be approximated by $E_{\vv{k}} \approx |f(\vv{k})|$, i.e., the nodal structure of a flat-band superconductor is dictated by the gap function. Note that, since $f$ is an analytic function, once it vanishes on an arbitrary small open subset in momentum space, it necessarily needs to vanish on the whole domain according to the identity theorem, i.e., the gap function needs to be zero on the whole domain. Hence, in two-dimensional superconductors, a surface node (or Bogoliubov Fermi surface) cannot be realized. Let us shift and rotate the coordinate system such that the analytic function~$f$ has a node at the origin and is not identically zero along the $k_1$ direction. The Weierstrass preparation theorem states that there exists a Weierstrass polynomial,
\begin{align}
    W(\mathbf{k})
    = k_1^{m} + a_{m-1}(k_2) k_1^{m-1} + \cdots + a_1(k_2) k_1 + a_0(k_2)
\end{align}
with \(m\ge 1\) and \(a_j(0)=0\) for \(j=0,1,\ldots,m-1\), such that the function \(f\) in a neighborhood of the origin admits the factorization,
\begin{align}
    f(\mathbf{k}) = u(\mathbf{k}) W(\mathbf{k}) = \Delta_T W(\vv{k}) [1 + o(1)]\,,
\end{align}
where $u(\mathbf{k})$ is analytic and nonvanishing near the origin~\cite{Weierstrass1895,Golubitsky1973,Krantz2002}. The degree $m$ is the first nonzero order of the directional derivative in the $k_1$-direction,
\begin{align}
    m = \mathrm{min}(j \ge 1; \partial_{k_1}^j f(\vv{k}) |_{\vv{k}=0} \neq 0) \,.
\end{align}
The set of zeros of $f(\vv{k})$ is completely determined by the functions $a_i$ and the degree $m$, so that the Weierstrass preparation theorem allows a classification of the nodal structure. In particular, the factorization of $W(\vv{k})$ into linear factors is in general given by
\begin{align}
    W(\mathbf{k}) &= \prod_{i=1}^I \big(k_1 - \lambda_i(k_2)\big)^{m_i} \,,
\end{align}
where $m = \sum_{i=1}^I m_i$ and the relation between the coefficients $a_i$ and $\lambda_i$ is given by the set of Vieta's formulas~\cite{viete1983aequationum}. The quasiparticle dispersion is then determined by the absolute value of $f(\vv{k})$,
\begin{align}
    E_{\vv{k}} = \Delta_T \prod_{i=1}^I \kle{\big(k_1 - \mathrm{Re}\,\lambda_i(k_2)\big)^2 + \big(\mathrm{Im}\,\lambda_i(k_2)\big)^2}^{m_i/2} \,.
\end{align}
As is clear from the factorization, the zero set of each linear factor with $\mathrm{Im}(\lambda_i) = 0$ produces a line node $k_1 - \mathrm{Re}\,\lambda_i(k_2)=0$ with shallowness $m_i$, while it becomes a point node of shallowness $m_i$ once $\mathrm{Im}(\lambda_i) \neq 0$. To distinguish both types, we also write the factorization as
\begin{align}
    \frac{E_{\vv{k}}}{\Delta_T} = \prod_{l=1}^L \big|k_1 - u_l(k_2)\big|^{q_l} \prod_{j = 1}^J \big((k_1 - v_j(k_2))^2 + w_j(k_2)^2\big)^{p_j/2},
    \label{eq:factorization-eigenvalue}
\end{align}
where $m = \sum_{l=1}^L q_l + \sum_{j=1}^J p_j$ and $w_{1,\hdots,J} \not\equiv 0$. Note that, even though $f$ is analytic, the functions $u_i,v_i,w_i$ defined in Eq.~\eqref{eq:factorization-eigenvalue} are not necessarily analytic. For example, the zero set of
\begin{align}
    f(\vv{k}) = k_1^2 - k_2^3 = (k_1 - k_2^{3/2}) (k_1 + k_2^{3/2})
    \label{eq:example-cusp}
\end{align}
exhibits a cusp at the origin. Nevertheless, the Newton-Puiseux theorem guarantees here that the functions $u_i,v_i,w_i$ admit convergent Puiseux expansions \cite{wall2004singular}. \\
\indent When $L = 0$ and $J \neq 0$, the gap function exhibits a point node, while the gap function produces at least one line node for $L\neq 0$ (with a pointlike enhancement at the origin if $J\neq 0$). In particular, if $L=1$, there is a single noncrossing line node, if $L=2$ and $u'_{l_1}(0)\neq u'_{l_2}(0)$ for $l_1\neq l_2$, the two line nodes cross transversely, if $L=2$ and $u'_{l_1}(0)=u'_{l_2}(0)$, they are tangent and form a double line, and if $L>2$, multiple line nodes meet potentially with higher-order tangencies. \\
\indent Thus, there are four types of nodal structures in superconducting systems with flat single-particle bands. We distinguish (i) fully gapped cases where $f(\vv{k})$ has no zeros, i.e., the Weierstrass preparation theorem does not apply and this case is therefore not discussed here, (ii) point nodes for $L=0$ and $J\neq 0$, (iii) single line nodes for $L=1$, and (iv) line-node crossings for $L\ge 2$, with ($J\neq 0$) or without ($J=0$) pointlike enhancement in the later two cases. For example, a gap function with dispersion $|f(\vv{k})| \propto |k_1|$ models a single line node of degree (or shallowness) $m = 1$ without pointlike enhancement and a gap function with dispersion $|f(\vv{k})| \propto |k_1^2 - k_2^2|^{m}$ models a line-node crossing of degree (or shallowness) $m$. When compared to the nomenclature of Ref.~\cite{mazidian2013anomalous}, the case of $m = 1$ ($m = 2$) corresponds to the crossing of two linear (shallow) line nodes. See also Ref.~\cite{chandrasekaran2020catastrophe} for more examples in a different context. \\
\indent Because the calculations become very complicated if one considers the general expression for the dispersion given in Eq.~\eqref{eq:factorization-eigenvalue}, in this work we restrict ourselves to the representative case with $v_{1,\hdots,J} \equiv 0$, $w_{1,\hdots,J}(k_2) = k_2$, and 
\begin{align}
    u_l(k_2) = \cot\klr{\frac{(2l + 1)\pi}{2L}}k_2 \,, \quad q_l = q\,.
\end{align}
Then, using polar coordinates, we find the dispersion to be given by
\begin{align}
    E_{\vv{k}} = \Delta_T k^m |\cos(L\theta)|^{q} \,,
    \label{eq:general-dispersion}
\end{align}
where $m = q L + \sum_{j=1}^J p_j$ is the total radial order of vanishing, $L$ is associated to the number of straight nodal lines through the origin, and $q$ is the shallowness per line, i.e., $E_{\vv{k}} \sim \Delta_T |t|^q$ near any of the nodal lines if local coordinates with $t$ transverse to the line are taken. Moreover, to simplify the calculations, we assume $J = 0$ if $L > 0$, so that the pointlike enhancement at the origin is absent and $m = qL$ for the line nodes. This is enough for our physical application later.
\section{Density of states}\label{sec:dos}
The most important quantity for the calculations in this work is the density of states (DOS), in particular, its low-energy behavior. In two dimensions, the DOS of the flat band with quasiparticle dispersion $E_{\vv{k}}$ is given by \cite{lapp2020experimental,hirobe2025anomalous}
\begin{align}
    D(E) = \int\!\frac{\mathrm{d}^2k}{(2\pi)^2} \delta(E - E_{\vv{k}}) \,.
    \label{eq:density-states}
\end{align}
The low-energy scaling laws corresponding to the cases of $L = 0$ (point nodes), $L = 1$ (line nodes without crossing), $L = 2$ (crossing of two line nodes), and $L > 2$ (crossing of three or more line nodes) are summarized in Table~\ref{tab:dos-scaling-laws}. 
\begin{table}[b!]
\caption{Low-energy scaling laws obtained for point and line nodes representable by the dispersion \eqref{eq:general-dispersion}. Here, $m > 0$ is the total order and $L > 0$ is associated to the number of crossing line nodes. Note that the order $m$ is always larger than or equal to the number of crossing line nodes.}\label{tab:dos-scaling-laws}
\renewcommand{\arraystretch}{1.4}
\begin{ruledtabular}
\begin{tabular}{lll}
  & node type & low-energy scaling \\ 
\colrule
    $L = 0$ & point nodes & $E^{2/m-1}$ \\
    $L = 1$ & line nodes without crossing & $E^{1/m-1}$ \\
    $L = 2$ & crossing of two line nodes & $E^{2/m-1} \ln(1/E)$ \\
    $L > 2$ & crossing of three or more line nodes & $E^{2/m-1}$
\end{tabular}
\end{ruledtabular}
\end{table}
\subsection{Point nodes}
The quasiparticle dispersion for a point node with order $m$ can be approximated by
\begin{align}
    E_{\vv{k}} \approx \Delta_T |\vv{k}|^{m} \,.
    \label{eq:quasi-energy-point-node}
\end{align}
We insert this dispersion relation into Eq.~\eqref{eq:density-states} to obtain
\begin{align}
    D(E) = \frac{4}{(2\pi)^2} \int_{[0,\infty]^2} \mathrm{d}^2k\,  \delta(E- \Delta_T(k_1^2 + k_2^2)^{m/2}) \,.
\end{align}
This integral can be calculated by introducing polar coordinates, which results in
\begin{align}
    D(E) = \frac{1}{2\pi m \Delta_T^{2/m}} E^{2/m - 1} \,.
\end{align}
\subsection{Line nodes}
Let us consider line nodes which correspond to the case $L > 0$. We insert the dispersion \eqref{eq:general-dispersion} into Eq.~\eqref{eq:density-states} to obtain
\begin{align}
    D(E) = \frac{1}{4\pi^2} \int_0^\infty \mathrm{d}k\, k \int_0^{2\pi} \mathrm{d}\theta\, \delta(E - \Delta_T k^m|\cos(L\theta)|^q) \,.
\end{align}
Since $L > 0$, this integral diverges if $q/m > 1/2$. Thus, we need to introduce a large momentum cutoff $\Lambda > 0$,
\begin{align}
    D(E) = \frac{1}{4\pi^2} \int_0^{2\pi} \mathrm{d}\theta \int_0^\Lambda \mathrm{d}k\,k\,\delta(E - \Delta_T k^m|\cos(L\theta)|^q) \,.
\end{align}
For a fixed $\theta$, the root is
\begin{align}
    k_0(\theta) = \kle{\frac{E}{\Delta_T |\cos(L\theta)|^q}}^{1/m} \,,
\end{align}
so that an evaluation of the $k$-integral gives
\begin{align}
    D(E) &= \frac{1}{4\pi^2m \Delta_T^{2/m}} E^{2/m - 1} \underbrace{\int_0^{2\pi} \mathrm{d}\theta\, \frac{\Theta(\Lambda - k_0(\theta))}{|\cos(L\theta)|^{2q/m}}}_{=\,I(E)} \,.
\end{align}
The remaining integral can be calculated by substituting $\phi = L\theta$ and by using the $\pi$-periodicity of $|\cos(\phi)|$ together with the symmetry of $|\cos(\phi)|$ about $\pi/2$. This gives
\begin{align}
    I(E) 
    &= 4\int_0^{\arccos((E/(\Delta_T \Lambda^m))^{1/q})} \mathrm{d}\phi \frac{1}{\cos^{2q/m}(\phi)} \,.
\end{align}
The incomplete beta function reads \cite[\S 8.39]{gradshteyn2014table} 
\begin{align}
    B_x(a,b) = \int_0^x\! t^{a-1} (1-t)^{b-1} \mathrm{d}t \equiv \frac{x^a}{a} {}_2F_1(a,1-b;1+a;x) \,,
\end{align}
which, for $x=1$, corresponds to the usual beta function. Therefore, we find the DOS to be given by
\begin{align}
    D(E) = \frac{E^{2/m-1}}{2\pi^2m \Delta_T^{2/m}}  B_{1-{(E/(\Delta_T \Lambda^m))^{2/q}}}( 1/2, 1/2 - q/m) \,,
\end{align}
and this integral is indeed non-existent for $q/m > 1/2$ as $\Lambda \to \infty$. Thus, we need to distinguish between three cases.

For $q/m<1/2$, the integral exists and no cutoff is needed. In the limit $\Lambda \to \infty$, we find the DOS to be given by
\begin{align}
    D(E) = \frac{B(\frac{1}{2},\frac{1}{2}-\frac{q}{m})}{2\pi^2m \Delta_T^{2/m}} E^{2/m-1} \,.
\end{align}

For $q/m = 1/2$, we make use of the fact that $B_x(1/2,0) = 2\mathrm{artanh}(\sqrt{x})$ \cite[\S 9.12]{gradshteyn2014table}. Since we have $\mathrm{artanh}(\sqrt{1 - x^2}) = \ln(2/x) + \mathcal{O}(x^{2})$ for $x \ll 1$, we find at leading order
\begin{align}
    D(E) 
    &= \frac{1}{\pi^2mq\Delta_T^{2/m}} E^{2/m-1} \ln(1/E) \,,
    \label{eq:dos-crossing-line-nodes-v1}
\end{align}
which also coincides with the result obtained in Ref.~\cite{hirobe2025anomalous}.

We proceed similarly for $q/m > 1/2$. By using the power series of the hypergeometric function \cite[\S 9.14]{gradshteyn2014table}, we find
\begin{align}
    B_{1-x}(a,b) = B(a,b) - \frac{x^b}{b} + \mathcal{O}(x^{b+1})
\end{align}
with $x = [E/(\Delta_T \Lambda^m)]^{2/q}$ in our case. Thus, we obtain the DOS as
\begin{align}
    D(E) 
    &= \frac{\Lambda^{2-m/q}}{\pi^2 (2q-m) \Delta_T^{1/q}} E^{1/q-1} \,. 
\end{align}
For instance, a line node with no crossings ($L = 1$) and no point-like enhancement has $q = m$ and a power law of $D(E) \propto E^{1/m-1}$.

\section{Low-temperature scaling laws}\label{sec:scaling-laws}
\subsection{Order parameter}\label{sec:order-parameter-laws}
First, let us discuss the temperature dependence of the order parameter. Since we assume that the bands are isolated, we can further approximate the self-consistent equation \eqref{eq:self-consistent-eq} by
\begin{align}
    d_\Gamma' \Delta_T \approx \frac{U_0}{2V} \sum_{\vv{k}} \frac{|f(\vv{k})|^2}{\Delta_T E_{\vv{k}}} [1 - 2n_{\mathrm{F}}(E_{\vv{k}})] \,, 
\end{align}
where we inserted $f(\vv{k}) = \Delta_T \sum_{i=1}^{d_\Gamma} \eta_i \psi_i^\Gamma(\vv{k})$, cf.\ Eq.~\eqref{eq:form-factor-definition}, and $d_\Gamma'$ represents the number of nonzero coefficients $\eta_i \neq 0$.
We further 
replace the momentum integral by an energy integral via Eq.~\eqref{eq:density-states}. The gap equation becomes
\begin{align}
    \Delta_T = \kle{\frac{U_0}{2d_\Gamma'} \int_0^\infty \mathrm{d}E\, D(E) E [1 - 2n_{\mathrm{F}}(E)]}^{1/2} \label{eq:energy-integrl-gap} \,.
\end{align}
According to Sec.~\ref{sec:dos}, the DOS of any nodal structure considered here has the form of
\begin{align}
    D(E) = \tilde{D}_0 \Delta_T^{-\alpha} E^{\alpha-1} \ln^\beta(1/E) \,,
\end{align}
where $\tilde{D}_0$ is a constant independent of temperature and energy, $\alpha > 0$ depends on the order $m$ (or shallowness $q$), and $\beta = 0,1$ depends on the number of crossing line nodes. We insert this form into Eq.~\eqref{eq:energy-integrl-gap} to obtain
\begin{align}
    \Delta_T = \kle{\frac{U_0 \tilde{D}_0}{2d_\Gamma'} \int_0^\infty \mathrm{d}E\, E^{\alpha} \ln^\beta(1/E) [1 - 2n_{\mathrm{F}}(E)]}^{1/(2 + \alpha)} \,.
\end{align}
The first term in the integrand is temperature-independent and corresponds to the order parameter at zero temperature $\Delta_0$. Therefore, the low-temperature scaling law of the difference $\Delta_0 - \Delta_T$ is determined by
\begin{align}
    \Delta_0 - \Delta_T = \frac{U_0\tilde{D}_0}{d_\Gamma' (\alpha+2)\Delta_0^{\alpha+1}}
    \int_0^\infty \mathrm{d}E\, \frac{E^\alpha \ln^\beta(1/E)}{e^{E/T} + 1} \,.
\end{align}
We substitute $x = E/T$ to obtain 
\begin{align}
    \Delta_0 - \Delta_T = C_\alpha T^{\alpha + 1} \ln^\beta(1/T) \quad (T \ll T_{\mathrm{c}})  \,,
\end{align}
where the proportionality constant $C_\alpha$ is given by
\begin{align}
    C_\alpha = \Gamma(\alpha + 1) \eta(\alpha + 1) \frac{U_0\tilde{D}_0}{d_\Gamma' (\alpha+2)\Delta_0^{\alpha+1}} \,.
    \label{eq:Calpha-constant}
\end{align}
\subsection{Superfluid weight}
The superfluid weight is defined by the second total derivative of the free energy with respect to the external gauge field $\vv{q} = \vv{A}$ as~\cite{hirschfeld1993effect,taylor2006pairing}
\begin{align}
    {D_{\mathrm{s},ij}(T) = \frac{1}{V}\frac{\mathrm{d}^2 F}{\mathrm{d}q_i\mathrm{d}q_j}\klsr{}_{\vv{q}=0}} \,.
\end{align}
We assume that the single-particle Hamiltonian is TRS and $N = -\partial\Omega/\partial\mu$ is constant in $\vv{q}$. Then, for unconventional pairings, it has been shown that the superfluid weight is given by \cite{lamponen2025superconductivity,buthenhoff2025functional}
\begin{align}
    D_{\mathrm{s},ij}(T) = D_{\mathrm{s},ij}^{\mathrm{conv}}(T) + D_{\mathrm{s},ij}^{\mathrm{geom}}(T) - D_{\mathrm{s},ij}^{\mathrm{func}}(T)\,.
    \label{eq:contributions-sw}
\end{align}
The first term is the conventional contribution and depends only on the curvature of the energy bands and the second term is the local part of the geometrical contribution. Analytical expressions for these contributions can be found, for example, in the Supplemental Material of Ref.~\cite{xie2020topology}. The third term represents the functional contribution (or the nonlocal part of the geometrical contribution) to the superfluid weight. For zero temperature, an analytical expression was derived in Ref.~\cite{buthenhoff2025functional}. Note that for conventional pairings, we can find a basis in which the functional contribution vanishes~\cite{huhtinen2022revisiting}. Since we consider flat-band superconductors, the conventional contribution vanishes, so that we only need to find the low-temperature scaling behaviors of the geometrical and functional contributions. Appendix~\ref{app:isolated-limit} collects explicit formulas for the superfluid weight in the isolated-band limit relevant to our discussion.
\subsubsection{Geometrical superfluid weight}
To analyze the low-temperature scaling law of the geometrical superfluid weight, we define
\begin{align}
    \Delta D_{\mathrm{s},ij}^{\mathrm{geom}} = D_{\mathrm{s},ij}^{\mathrm{geom}}(0) - D_{\mathrm{s},ij}^{\mathrm{geom}}(T) 
\end{align}
and take the derivative with respect to the temperature \cite{hirobe2025anomalous}. According to the chain rule, we have
\begin{align}
    \frac{\mathrm{d}(\Delta D_{\mathrm{s},ij}^{\mathrm{geom}})}{\mathrm{d}T} = \frac{\partial (\Delta D_{\mathrm{s},ij}^{\mathrm{geom}})}{\partial T} + \frac{\partial (\Delta D_{\mathrm{s},ij}^{\mathrm{geom}})}{\partial \Delta_T} \frac{\mathrm{d} \Delta_T}{\mathrm{d} T} \,.
\end{align}
Here, functional derivatives are not necessary because the $\vv{k}$-dependent part of the gap function is not temperature dependent. Note that the temperature dependence of the order parameter was ignored in Ref.~\cite{hirobe2025anomalous}. Let us first consider the first contribution. In the isolated flat-band limit, the geometrical superfluid weight is provided by Eq.~\eqref{eq:geom-weight-isolated-flat-bands},
\begin{align}
    \Delta D_{\mathrm{s},ij}^{\mathrm{geom}} \approx \frac{8}{V} \sum_{\vv{k}} \frac{n_{\mathrm{F}}(E_{\vv{k}})}{E_{\vv{k}}} |f(\vv{k})|^2 g_{ij}(\vv{k})\,,
\end{align}
where $g_{ij}(\vv{k})$ is the quantum metric of the flat band \cite{provost1980riemannian}. We calculate the derivative with respect to the temperature in the continuous limit to obtain
\begin{align}
    \frac{\partial (\Delta D_{\mathrm{s},ij}^{\mathrm{geom}})}{\partial T} = \frac{8}{T^2} \int\! \frac{\mathrm{d}^2k}{(2\pi)^2} g_{ij}(\vv{k})  \frac{e^{E_{\vv{k}}/T}E_{\vv{k}}^2}{(e^{E_{\vv{k}}/T} + 1)^2} \,,
\end{align}
where we used $|f(\vv{k})|^2 = E_{\vv{k}}^2$ for a flat band. Analogously to Refs.~\cite{lapp2020experimental,hirobe2025anomalous}, we insert the DOS and replace the momentum integral by an energy integral as
\begin{align}
    \frac{\partial (\Delta D_{\mathrm{s},ij}^{\mathrm{geom}})}{\partial T} = 8\int_0^\infty \mathrm{d}E\, D(E) \klr{\frac{E}{T}}^2 \frac{e^{E/T}}{(e^{E/T} + 1)^2} \kla{g_{ij}}_E\,,
    \label{eq:temp-dependence-first-term-exp-value}
\end{align}
where $\kla{\cdot}_E$ is defined at a given energy $E$ by
\begin{align}
    \kla{g_{ij}}_E = \frac{1}{D(E)} \int\! \frac{\mathrm{d}^2k}{(2\pi)^2} \delta(E - E_{\vv{k}}) g_{ij}(\vv{k}) \,.
\end{align}
To calculate this expectation value, we proceed similarly to Ref.~\cite{hirobe2025anomalous}. In particular, we expand $g_{ij}(\vv{k})$ around the point node which is set to the origin without loss of generality,
\begin{align}
    g_{ij}(\vv{k}) = \kla{g_{ij}}_{E=0} + \mathcal{O}(|\vv{k}|) \,,
\end{align}
so that we obtain
\begin{align}
    \kla{g_{ij}}_E = \kla{g_{ij}}_{E=0} + \mathcal{O}(E^{\gamma}) 
\end{align}
with some exponent $\gamma > 0$ depending on the node type. We insert this result into Eq.~\eqref{eq:temp-dependence-first-term-exp-value} and substitute $x = E/T$. The leading order is then provided by
\begin{align}
    \frac{\partial (\Delta D^{\mathrm{geom}}_{\mathrm{s},ij})}{\partial T} = 8\kla{g_{ij}}_{E=0} \int_0^\infty \mathrm{d}x \frac{x^2 e^x}{(e^x + 1)^2} D(xT)T \,.
    \label{eq:geom-first-term}
\end{align}
Similarly, the leading order of the derivative with respect to $\Delta_T$ is given by
\begin{align}
    \frac{\partial (\Delta D_{\mathrm{s},ij}^{\mathrm{geom}})}{\partial \Delta_T} = \frac{8\kla{g_{ij}}_{E=0}}{\Delta_0} \int_0^\infty \mathrm{d}x\, \frac{x[e^{x} (1 - x) + 1]}{(e^{x} + 1)^2} D(xT) T^2 \,.
    \label{eq:geom-second-term}
\end{align} 
For a node with the low-energy scaling law of $D(E) = D_0 E^{\alpha-1}$, where $D_0$ is the energy-independent constant depending on the nodal structure, we find
\begin{align}
    \frac{\partial (\Delta D_{\mathrm{s},ij}^{\mathrm{geom}})}{\partial T} 
    &= 8D_0 \kla{g_{ij}}_{E=0} \Gamma(\alpha + 2) \eta(\alpha + 1) T^{\alpha} \,.
\end{align}
Similarly, we obtain 
\begin{align}
    \frac{\partial (\Delta D_{\mathrm{s},ij}^{\mathrm{geom}})}{\partial \Delta_T} 
    &= -\frac{8D_0 \kla{g_{ij}}_{E=0} \alpha \Gamma(\alpha + 1) \eta(\alpha + 1)}{\Delta_0} T^{\alpha + 1} \,.
\end{align}
In particular, by using the results obtained in Sec.~\ref{sec:order-parameter-laws}, we further find
\begin{align}
    \frac{\partial (\Delta D_{\mathrm{s},ij}^{\mathrm{geom}})}{\partial \Delta_T} \frac{\mathrm{d} \Delta_T}{\mathrm{d} T} \propto T^{2\alpha + 1}\,,
\end{align}
where the proportionality constant depends on the nodal structure. Therefore, the temperature dependence due to the order parameter is subleading and we conclude the low-temperature scaling law of
\begin{align}
    \Delta D_{\mathrm{s},ij}^{\mathrm{geom}} \propto T^{\alpha + 1} \quad (T \ll T_{\mathrm{c}}) 
\end{align}
for the geometrical superfluid weight. For example, a point node or a crossing of more than two line nodes has the exponent $\alpha = 2/m$. On the other hand, a line node without crossing has the exponent $\alpha = 1/m$. \\
\indent We then suppose that the low-energy scaling law of the DOS is given by $D(E) = D_0 E^{\alpha - 1} \ln(1/E)$. This case for $\alpha = 2/m$ corresponds to a crossing of two line nodes. Because of $\ln[1/(xT)] = \ln(1/T) + \ln(1/x)$, we can divide the integrals in Eq.~\eqref{eq:geom-first-term} and Eq.~\eqref{eq:geom-second-term} into two terms each. The evaluation of the integrals containing $\ln(1/T)$ is identical to the previous case. Moreover, the second terms containing $\ln(1/x)$ are subleading because the integrals,
\begin{align}
    \int_0^\infty \mathrm{d}x \frac{e^x x^{\alpha + 1} \ln(1/x)}{(e^x + 1)^2} &< \infty \,,\\
    \int_0^\infty \mathrm{d}x \frac{\ln(1/x)x^\alpha[e^x(1 - x) + 1]}{(e^x + 1)^2} &< \infty \,,
\end{align}
both exist for $\alpha > 0$. Thus, the low-temperature scaling law is given by
\begin{align}
    \Delta D_{\mathrm{s},ij}^{\mathrm{geom}} \propto T^{\alpha + 1} \ln(1/T) \quad (T \ll T_{\mathrm{c}}) \,.
\end{align}
\subsubsection{Functional superfluid weight}
Analogously to the geometrical superfluid weight, we define
\begin{align}
    \Delta D_{\mathrm{s},ij}^{\mathrm{func}} = D_{\mathrm{s},ij}^{\mathrm{func}}(0) - D_{\mathrm{s},ij}^{\mathrm{func}}(T)\,.
\end{align}
Under the assumption of isolated bands, the functional superfluid weight is provided by Eq.~\eqref{eq:func-weight-isolated-flat-bands}, 

\begin{align}
    \Delta D_{\mathrm{s},ij}^{\mathrm{func}} \approx& \frac{4}{V} \sum_{\vv{k},\vv{k}'} \sum_{\alpha,\beta = 1}^{N_\mathrm{B}} \sum_{\mu,\nu \in \klg{\mathrm{R},\mathrm{I}}}  M_{\alpha\mu,\beta\nu}^{-1}(\vv{k},\vv{k}')  f^{\bar{\mu}}(\vv{k}) f^{\bar{\nu}}(\vv{k}')  \nonumber \\
    &\times {s_\mu s_\nu} \frac{{2(n_{\mathrm{F}}(E_{\vv{k}}) + n_{\mathrm{F}}(E_{\vv{k}'}))} {-4}n_{\mathrm{F}}(E_{\vv{k}}) n_{\mathrm{F}}(E_{\vv{k}'}) }{E_{\vv{k}}E_{\vv{k}'}} \nonumber \\
    &\times h_{ij,\alpha\beta}(\vv{k},\vv{k}'),
\end{align}
where we define $s_{\mathrm{R},\mathrm{I}} = \pm 1$, $f^{\mathrm{R}} = \mathrm{Re}(f)$ and $f^{\mathrm{I}} = \mathrm{Im}(f)$ indicate the real and imaginary parts, $\bar{\mu} = \mathrm{R}$ if $\mu = \mathrm{I}$ and $\bar{\mu} = \mathrm{I}$ if $\mu = \mathrm{R}$, $h_{ij,\alpha\beta}(\vv{k},\vv{k}')$ is a nonlocal multi state quantum geometric quantity defined in Eq.~\eqref{eq:multi-state-quantum-geometric-term}, and the components of the matrix $M^{-1}$ are
given by
\begin{align}
    M^{-1}_{\alpha\mu,\beta\nu}(\vv{k},\vv{k}') =& \frac{1}{V}\! \klg{{\mathrm{Re}[U(\vv{k},\vv{k}')] \delta_{\mu\nu} - \mathrm{Im}[U(\vv{k},\vv{k}')] \epsilon_{\mu\nu}}}\! \delta_{\alpha\beta} \nonumber \\
    &+ \mathcal{O}(T^\gamma)
\end{align}
with some exponent $\gamma > 0$ depending on the node type and determined by Eq.~\eqref{eq:pi-def}. Hence, at leading order in temperature, the functional contribution to the superfluid weight is given by
\begin{align}
    \Delta D_{\mathrm{s},ij}^{\mathrm{func}} \approx& \frac{4}{V^2} \sum_{\vv{k},\vv{k}'}\kle{2(n_{\mathrm{F}}(E_{\vv{k}}) + n_{\mathrm{F}}(E_{\vv{k}'})) - 4 n_{\mathrm{F}}(E_{\vv{k}}) n_{\mathrm{F}}(E_{\vv{k}'})} \nonumber \\
    &\times \mathrm{Re}\!\kle{U(\vv{k},\vv{k}') e^{i[\phi(\vv{k}') - \phi(\vv{k})]}} h_{ij}(\vv{k},\vv{k}') \,,
\end{align}
where $\phi(\vv{k}) = \mathrm{arg}(f(\vv{k}))$ and $h_{ij} \equiv \sum_{\alpha} h_{ij,\alpha\alpha}$. We take the derivative with respect to the temperature to obtain
\begin{align}
    \frac{\mathrm{d}(\Delta D_{\mathrm{s},ij}^{\mathrm{func}})}{\mathrm{d}T} = \frac{\partial (\Delta D_{\mathrm{s},ij}^{\mathrm{func}})}{\partial T} + \frac{\partial (\Delta D_{\mathrm{s},ij}^{\mathrm{func}})}{\partial \Delta_T} \frac{\mathrm{d} \Delta_T}{\mathrm{d} T} \,.
\end{align}
The first term is given by 
\begin{widetext}
 \begin{align}
    \frac{\partial (\Delta D_{\mathrm{s},ij}^{\mathrm{func}})}{\partial T} = \frac{8}{V^2} \sum_{\vv{k},\vv{k}'}  \frac{1}{T^2} \frac{E_{\vv{k}} e^{E_{\vv{k}}/T} ( e^{2E_{\vv{k}'}/T} - 1)  + E_{\vv{k}'}e^{E_{\vv{k}'}/T} (e^{2E_{\vv{k}}/T} - 1) }{(e^{E_{\vv{k}}/T} + 1)^2 (e^{E_{\vv{k}'}/T} + 1)^2} {\mathrm{Re}\!\kle{U(\vv{k},\vv{k}') e^{i[\phi(\vv{k}') - \phi(\vv{k})]}}} h_{ij}(\vv{k},\vv{k}') \,.
\end{align}   
We replace both the momentum sums by the energy integrals to obtain 
\begin{align}
    \frac{\partial (\Delta D_{\mathrm{s},ij}^{\mathrm{func}})}{\partial T} =  8\int_{E,E'=0}^\infty \frac{D(E)D(E')}{T^2} \frac{E e^{E/T}(e^{2E'/T} - 1) + E' e^{E'/T} (e^{2E/T} - 1) }{(e^{E/T} + 1)^2 (e^{E'/T} + 1)^2} \kla{{\mathrm{Re}\!\kle{U(\vv{k},\vv{k}') e^{i[\phi(\vv{k}') - \phi(\vv{k})]}}}h_{ij}(\vv{k},\vv{k}')}_{E,E'},
    \label{eq:zwischen-erg}
\end{align}
where 
\begin{align}
    \kla{F}_{E,E'} &= \frac{1}{D(E)D(E')} \sum_{\vv{k},\vv{k}'} \delta(E - E_{\vv{k}}) \delta(E' - E_{\vv{k}'}) F(\vv{k},\vv{k}') \,. 
\end{align}
Since the basis functions are analytic functions, the pairing potential admits the Taylor expansion,
\begin{align}
    U(\vv{k},\vv{k}') = U_0 \sum_{i=1}^{d_\Gamma} \sum_{M,N \ge 0} \frac{1}{M!N!} \klr{\partial^M \bar{\psi}_i^\Gamma(0)} \klr{\partial^N \psi_i^\Gamma(0)} \vv{k}^M (\vv{k}')^N\,,
\end{align}
where $M,N \in \mathbb{N}^2_0$ are multi-indices. We then denote the lowest non-vanishing order of $U(\vv{k},\vv{k}')$ by $M_0,N_0$ and define the superconducting state dependent exponent $b \ge 0$ via the equation,
\begin{align}
    \frac{U_0}{M_0!N_0!} \sum_{i=1}^{d_\Gamma} \mathrm{Re}\!\kle{\klr{\partial^{M_0} \bar{\psi}_i^\Gamma(0)} \klr{\partial^{N_0} \psi_i^\Gamma(0)} \Big\langle e^{i[\phi(\vv{k}') - \phi(\vv{k})]}\vv{k}^{M_0} (\vv{k}')^{N_0}\Big\rangle_{E,E'}} = c (EE')^b \,,
    \label{eq:gamma-exponent-definition}
\end{align}
where the proportionality coefficient $c$ depends on the order parameter at zero temperature. According to Eq.~\eqref{eq:pairing-potential}, we always have $b=1$ for one-dimensional superconducting states with $d_\Gamma = 1$. Moreover, the same holds for all two-dimensional cases considered below in Sec.~\ref{sec:application}.
We insert this expansion and substitute $x = E/T$ and $y = E'/T$ in Eq.~\eqref{eq:zwischen-erg},
\begin{align}
    \frac{\partial (\Delta D_{\mathrm{s},ij}^{\mathrm{func}})}{\partial T} = 8c \kla{h_{ij}}_{E,E'=0} \int_{x,y=0}^\infty \frac{x^{{b}} y^{{b}} [xe^x (e^{2y} - 1) + y e^y (e^{2x} - 1)]}{(e^{x} + 1)^2 (e^{y} + 1)^2} D(xT) D(yT) T^{{1 + 2b}} \,,
\end{align}
\end{widetext}
and then insert $D(E) = D_0 E^{\alpha - 1} \ln^\beta(1/E)$. Since these integrals are divergent, we introduce an energy cutoff $\Lambda > 0$ and find the leading order to be given by 
\begin{align}
    \frac{\partial (\Delta D_{\mathrm{s},ij}^{\mathrm{func}})}{\partial T} = 16D_0^2 c \kla{h_{ij}}_{E,E'=0} K_{\alpha\beta b} T^{{\alpha + b - 1}} \ln^{\beta}(1/T),
\end{align}
where the Feynman trick and the fact that the incomplete Gamma functions obey the property $\Gamma(n,x) + \gamma(n,x) = \Gamma(n)$, cf.\ \cite[\S 8.35]{gradshteyn2014table}, is used to obtain
\begin{align}
    K_{\alpha\beta b} = (-1)^\beta \eta(\alpha + b) \Gamma(\alpha + b + 1) \frac{\mathrm{d}^\beta}{\mathrm{d}s^\beta}\klsr{}_{s=0} \frac{\Lambda^{\alpha + b - s}}{\alpha + b - s} \,.
\end{align}
Similarly, the derivative with respect to $\Delta_T$ is provided by
\begin{align}
    \frac{\partial(\Delta D_{\mathrm{s},ij}^{\mathrm{func}})}{\partial \Delta_T} =& -\frac{16c\kla{h_{ij}}_{E,E'=0}}{\Delta_0} \int_{x,y=0}^\infty \frac{x^{{b}} y^{{b}} [xe^x(e^y - 1)]}{(e^x + 1)^2 (e^y + 1)} \nonumber \\
    &\times D(xT) D(yT) T^{{2 + 2b}} .
\end{align}
Again, we insert the DOS of $D(E) = D_0 E^{\alpha - 1} \ln^\beta(1/E)$ to obtain
\begin{align}
    \frac{\partial(\Delta D_{\mathrm{s},ij}^{\mathrm{func}})}{\partial \Delta_T} \frac{\mathrm{d} \Delta_T}{\mathrm{d}T} =& \frac{16D_0^2 c \kla{h_{ij}}_{E,E'=0} K_{\alpha\beta b}}{\Delta_0} \nonumber \\
    &\times T^{{2\alpha + b}} \ln^{{2\beta}}(1/T)\,,
\end{align}
Therefore, we conclude that the low-temperature scaling law of the functional superfluid weight is given by
\begin{align}
    \Delta D_{\mathrm{s},ij}^{\mathrm{func}} \propto T^{{\alpha + b}} \ln^{{\beta}}(1/T) \quad (T\ll T_{\mathrm{c}}) \,.
\end{align}
\subsection{Tunneling conductance}
The tunneling current between a superconductor and a normal conductor can be phenomenologically described by \cite[\S 3.8]{tinkham2004introduction}
\begin{align}
    I_{sn} = A \int_{-\infty}^\infty \mathrm{d}E D(E) \kle{n_{\mathrm{F}}(E) - n_{\mathrm{F}}(E + V)} \,,
\end{align}
where $V$ is the bias voltage and we work with the convention $e = 1$ as indicated below Eq.~\eqref{eq:grand-potential-mf-bcs}. The proportionality coefficient $A$ depends on the tunneling matrix elements and the DOS of the normal conductor, both of which are assumed to be energy independent.
This expression corresponds to the tunneling (low transparency) limit in which transport is characterized by the quasiparticle DOS. If the interface transparency is not small, or interface-induced bound states are relevant, Andreev reflection becomes important and the DOS-based description may no longer be sufficient. In that case, a treatment within the Blonder-Tinkham-Klapwijk (BTK) framework is more appropriate \cite{blonder1982transition,zhu2016bogoliubov}. \\
\indent The linear conductance is given by the derivative of the tunneling current with respect to the bias voltage at $V = 0$,
\begin{align}
    G_{sn} 
    &= A \int_{-\infty}^\infty \mathrm{d}x\, \frac{e^{x}}{(e^{x} + 1)^2} D(xT) \,.
\end{align}
We then insert $D(E) = D_0 E^{\alpha - 1} \ln^\beta(1/E)$ to obtain at leading order
\begin{align}
    G_{sn} 
    &= 2AD_0 \Gamma(\alpha) \eta(\alpha - 1) T^{\alpha - 1} \ln^\beta(1/T) \,,
\end{align}
which is the low-temperature scaling law of the tunneling conductance.
\subsection{Specific heat and Sommerfeld coefficient}
For low temperatures, the specific heat $C$ is given by \cite{sigrist1991phenomenological,lapp2020experimental}
\begin{align}
    C &= \int_0^\infty \mathrm{d}E\, \kle{-D(E) \frac{E^2}{T}  \frac{\mathrm{d}n_{\mathrm{F}}(E)}{\mathrm{d}E} + \frac{\partial D(E)}{\partial \Delta_T} \frac{\mathrm{d}\Delta_T}{\mathrm{d}T} E n_{\mathrm{F}}(E)} \,,
\end{align}
where the first term is due to the temperature dependence of the Fermi distribution function and the second term is due to the temperature dependence of the order parameter. Since $\Delta_0 - \Delta_T = C_\alpha T^{\alpha + 1} \ln^\beta(1/T)$ for $D(E) = \tilde{D}_0 \Delta_T^{-\alpha} E^{\alpha - 1} \ln^\beta(1/E)$ according to Sec.~\ref{sec:order-parameter-laws}, the leading order of $\partial D(E)/\partial \Delta_T$ is constant in temperature. Therefore, if we insert the DOS and substitute $x = E/T$, we obtain at leading order
\begin{align}
    C =& \frac{\tilde{D}_0}{\Delta_0^\alpha} \bigg[ \int_0^\infty \mathrm{d}x\, \frac{x^{\alpha + 1}  e^x}{(e^x + 1)^2} T^{\alpha} \ln^\beta(1/T) \nonumber \\
    &+ \frac{C_\alpha \alpha(\alpha + 1)}{\Delta_0} T^{2\alpha+1} \ln^{2\beta}(1/T) \int_0^\infty \mathrm{d}x\, \frac{x^\alpha}{e^x + 1} \bigg] \,,
\end{align}
where the constant $C_\alpha$ is defined in Eq.~\eqref{eq:Calpha-constant}.
We can clearly see that the first term dominates and thus
\begin{align}
    C \propto T^\alpha \ln^\beta(1/T) \quad (T \ll T_{\mathrm{c}}) \,.
\end{align}
Using this result, we determine the Sommerfeld coefficient $\gamma = C/T$ following the scaling law of
\begin{align}
    \gamma \propto T^{\alpha - 1} \ln^\beta(1/T) \quad (T \ll T_{\mathrm{c}}) \,.
\end{align}
\subsection{NMR spin-lattice relaxation rate}
The NMR spin-lattice relaxation rate is defined as \cite{sigrist1991phenomenological,fernandes2011scaling,lapp2020experimental}
\begin{align}
    \frac{1}{T_1T} = -\beta_{\mathrm{NMR}} \int_0^\infty \mathrm{d}E\, D^2(E) \frac{\mathrm{d}n_{\mathrm{F}}(E)}{\mathrm{d}E} \,,
\end{align}
where $\beta_{\mathrm{NMR}}$ is a constant containing the normal-state relaxation rate. We then insert $D(E) = D_0 E^{\alpha - 1} \ln^\beta(1/E)$ to obtain the low-temperature scaling law of
\begin{align}
    \frac{1}{T_1T} \propto T^{2\alpha - 2} \ln^{2\beta}(1/T) \,.
\end{align}
\section{Application to $C_{6v}$-symmetric systems}\label{sec:application}
\begin{table*}[t!]
\caption{Selection of possible unconventional superconducting states for a system with $C_{6v}$ symmetry and the corresponding low-temperature scaling laws of the order parameter $\Delta_T$, the superfluid weight $D_{\mathrm{s}}$, the tunneling conductance $G_{sn}$, the specific heat $C$, the Sommerfeld coefficient $\gamma$, and the NMR spin-lattice relaxation rate $1/(T_1T)$. 
}\label{tab:c6v-scaling-laws}
\renewcommand{\arraystretch}{1.4}
\begin{ruledtabular}
\begin{tabular}{llllllll}
 irrep & basis function & $\Delta_T$ & $D_{\mathrm{s}}$ & $G_{sn}$ & $C$ & $\gamma$ & $1/(T_1T)$ \\ 
\colrule
     $A_1$ & $k_1^2 + k_2^2$ & $T^2$ & $T^2$ & $\mathrm{const}$ & $T^2$ & $\mathrm{const}$ & $\mathrm{const}$ \\
     & $k_1^2(k_1^2 - 3k_2^2)^2 - k_2^2(3k_1^2 - k_2^2)^2$ & $T^{4/3}$ & $T^{4/3}$ & $T^{-2/3}$ & $T^{1/3}$ & $T^{-2/3}$ & $T^{-4/3}$ \\
     $A_2$ & $k_1k_2(k_1^2 - 3k_2^2)(k_2^2 - 3k_1^2)$ & $T^{4/3}$ & $T^{4/3}$ & $T^{-2/3}$ & $T^{1/3}$ & $T^{-2/3}$ & $T^{-4/3}$ \\
      $B_1$ & $k_1(k_1^2 - 3k_2^2)$ & $T^{5/3}$ & $T^{5/3}$ & $T^{-1/3}$ & $T^{2/3}$ & $T^{-1/3}$ & $T^{-2/3}$ \\
      $B_2$ & $k_2(3k_1^2 - k_2^2)$ & $T^{5/3}$ & $T^{5/3}$ & $T^{-1/3}$ & $T^{2/3}$ & $T^{-1/3}$ & $T^{-2/3}$  \\
      $E_1$ & $k_1, k_2$ & $T^2$ & $T^2$ & $\mathrm{const}$ & $T^2$ & $\mathrm{const}$ & $\mathrm{const}$ \\
      & $k_1 \pm ik_2$ & $T^3$ & $T^3$ & $T $ & $T^2$ & $T$ & $T^2$ \\
     $E_2$ & $k_1^2 - k_2^2, k_1k_2$ & $T^2\ln(1/T)$ & $T^2\ln(1/T)$ & $\ln(1/T)$ & $T\ln(1/T)$ & $\ln(1/T)$ & $\ln^2(1/T)$ \\
      & $k_1^2 - k_2^2 \pm 2i k_1k_2$ & $T^2$ & $T^2$ & $\mathrm{const}$ & $T^2$ & $\mathrm{const}$ & $\mathrm{const}$ 
\end{tabular}
\end{ruledtabular}
\end{table*}
In 2018, flat-band superconductivity was observed for the first time in the magic-angle twisted bilayer graphene (MATBG), where two graphene layers twisted by $1.1^\circ$ host nearly flat moiré bands  \cite{cao2018correlated,cao2018unconventional}. Since then, related observations have followed in twisted trilayer graphene \cite{park2021tunable,hao2021electric}, twisted bilayer tungsten diselenide \cite{xia2025superconductivity,guo2025superconductivity}, and layered kagome metals \cite{ortiz2020cs,ortiz2021superconductivity}. Many of these systems realize hexagonal settings. In particular, the low-energy moiré bands of ideal magic-angle twisted bilayer graphene are widely modeled to respect an emergent $D_6 \simeq C_{6v}$ symmetry \cite{angeli2018emergent,zou2018band}. As a specific representative example, we shall apply our derived scaling laws to flat-band superconductors with $C_{6v}$ symmetry. \\
\indent The hexagonal group $C_{6v}$ possesses six irreducible representations, where three of them are even-parity representations (spin-singlet states) and the other three are odd-parity representations (spin-triplet states) \cite{powell2010symmetry,black2014chiral}. Table~\ref{tab:c6v-scaling-laws} presents the basis function of each irreducible representation together with the corresponding low-temperature scaling laws. For all superconducting states considered in Table~\ref{tab:c6v-scaling-laws}, we have $b=1$, and the other parameters $m$, $L$, and $q$ can be found such that the basis functions can be cast into the dispersion given in Eq.~\eqref{eq:general-dispersion}. \\ 
\indent In Ref.~\cite{tanaka2024superfluid}, the low-temperature behavior of the superfluid weight in the magic-angle twisted bilayer graphene was experimentally measured. The experimental data were fitted to a power-law scaling $T^n$, where $n \simeq 2.08$ (hole-doped), $n \simeq 2.44$ (electron-doped), and $n \in [2,3]$ across the dome were reported. Since higher-order corrections in the low-temperature scaling of the superfluid weight can bias a power-law fit, we expect the presence of a superconducting state with a scaling exponent of $n \simeq 2$. When compared to Table~\ref{tab:c6v-scaling-laws}, this indicates the possibility of an extended $s$-wave, nematic $p$-wave, or chiral $d_{x^2-y^2} + id_{xy}$-wave state. Note that a nematic $d$-wave state would tend to produce an effective scaling exponent of $n < 2$. Due to the fact that the constant function is another element of the $A_1$ representation, the point node of the extended $s$-wave state is not stable and, hence, unlikely to be present in MATBG. Moreover, experiments in MATBG indicate the superconducting order parameter to be non-chiral \cite{cao2021nematicity}, which excludes the possibility of the chiral $d_{x^2-y^2} + id_{xy}$-wave state. Accordingly, we are left with the expectation of the nematic $p$-wave state in MATBG. This finding is consistent with Refs.~\cite{Sboychakov2024coexistence,kagan2025anomalous}, where nematic superconductivity similar to a $p$-wave order is anticipated in MATBG. Nevertheless, it is important to note that the scaling laws obtained in Table~\ref{tab:main-result} and Table~\ref{tab:c6v-scaling-laws} may change when the concentration of strongly scattering impurities is nonzero, similarly to Ref.~\cite{hirschfeld1993effect}.

\section{Summary and discussion}
We calculated the low-temperature scaling laws of the order parameter, the superfluid weight, the tunneling conductance, the specific heat, the Sommerfeld coefficient, and the spin-lattice relaxation rate in flat-band superconductors with unconventional pairing. The results obtained for different node types are collected in Table~\ref{tab:main-result}. In particular, by comparing the scaling laws of the geometrical and functional contributions to the superfluid weight, we found them to be of the same order for all one-dimensional superconducting states and most of the higher-dimensional superconducting states and conclude that the scaling laws of the superfluid weight reported in Ref.~\cite{hirobe2025anomalous} are intact. However, once we have a superconducting state with $b < 1$, the functional contribution might become dominant such that the scaling laws in Ref.~\cite{hirobe2025anomalous} need to be adjusted.  \\
\indent Several generalizations of our work are possible. First of all, we mostly considered the dispersion given in Eq.~\eqref{eq:general-dispersion} which is a special case of the dispersion in Eq.~\eqref{eq:factorization-eigenvalue}. It would be then interesting to further explore nodal structures not covered here but still allowed by the Weierstrass preparation theorem. For example, node types that exhibit cusps such as the dispersion in Eq.~\eqref{eq:example-cusp} may appear in non-centrosymmetric crystals \cite{bauer2012non}, and probably give rise to different exponents or additional logarithms in the scaling laws via the Puiseux expansion. Also, it remains to be clarified to what extent these scaling laws carry over to quasicrystals \cite{sun2025geometric}, and how the temperature dependence changes near the critical temperature \cite{penttila2025flat}. \\
\indent Lastly, we considered a selection of possible superconducting states (extended $s$-wave, chiral and nematic $p$-wave, $d$-wave, etc.)~that may appear in systems with $C_{6v}$ symmetry such as the magic-angle twisted bilayer graphene. The scaling laws obtained here are presented in Table~\ref{tab:c6v-scaling-laws}. Under the assumption that the superconducting state of MATBG is non-chiral (cf.\ Ref.~\cite{cao2021nematicity}), the comparison of our predictions in Table~\ref{tab:c6v-scaling-laws} with the experimental measurements of Ref.~\cite{tanaka2024superfluid} for the magic-angle twisted bilayer graphene indicates that nematic $p$-wave superconductivity is the most likely possibility. We believe that further low-temperature experiments in hexagonal moiré systems would be instructive to test the predicted scaling laws and to quantify to what extent the scaling relations can be used as a diagnostic of the pairing symmetry and possible topological phases.

\begin{acknowledgments}
This work was supported by the doctoral scholarship program of the German Academic Scholarship Foundation (M.B.) and by JSPS KAKENHI Grant No.~JP21K03384 (Y.N.). 
\end{acknowledgments}

\section*{Data availability}
No data were created or analyzed in this study.

\appendix

\onecolumngrid
\section{Superfluid weight of isolated flat-band superconductors for arbitrary temperature}\label{app:isolated-limit}
In this appendix, we derive expressions for the superfluid weight in the isolated flat-band limit necessary for the discussion of the low-temperature behavior. While Ref.~\cite{buthenhoff2025functional} derived the expressions only in the zero-temperature limit and assumed the presence of a real-valued pairing potential, here we derive the expressions for the superfluid weight of an isolated flat-band superconductor with a complex-valued pairing potential at arbitrary temperature. We follow the derivation presented in Ref.~\cite{buthenhoff2025functional}. Throughout this calculation, we assume that the gap function is proportional to the identity matrix (uniform pairing condition) and that the band with index $n_0$ is flat and located at the chemical potential. As stated in Eq.~\eqref{eq:contributions-sw}, the superfluid weight consists of three contributions. These can be obtained via derivatives of the mean-field BCS grand potential $\Omega$, defined in Eq.~\eqref{eq:grand-potential-mf-bcs}, via 
\begin{align}
    D_{\mathrm{s},ij}^{\mathrm{conv}}(T) + D_{\mathrm{s},ij}^{\mathrm{geom}}(T) = \frac{1}{V}\frac{\partial^2 \Omega}{\partial q_i \partial q_j} \klsr{}_{\vv{q}=0}\,, \qquad D_{\mathrm{s},ij}^{\mathrm{func}}(T) &= \frac{1}{V} \sum_{\vv{k},\vv{k}'} S_{i,\alpha\mu}(\vv{k}) M^{-1}_{\alpha\mu,\beta\nu}(\vv{k},\vv{k}') S_{j,\beta\nu}(\vv{k}')\,,
\end{align}
where
\begin{align}
    S_{i\alpha\mu}(\vv{k}) &\coloneqq \frac{\delta}{\delta f^\mu_{\alpha}(\vv{q};\vv{k})}\klr{\frac{\partial\Omega}{\partial q_i}} \klsr{}_{\vv{q}=0}\,, \qquad 
    M_{\alpha\mu,\beta\nu}(\vv{k},\vv{k}') \coloneqq \frac{\delta^2 \Omega}{\delta f^\mu_{\alpha}(\vv{q};\vv{k}) \delta f^\nu_{\beta}(\vv{q};\vv{k}')} \klsr{}_{\vv{q}=0} \,,
\end{align}
and $f^{\mathrm{R}} = \mathrm{Re}(f)$ and $f^{\mathrm{I}} = \mathrm{Im}(f)$ indicate the real and imaginary parts. In the case of isolated bands, due to TRS, the Bogoliubov eigenvalues of the $n$th band are given by
\begin{align}
    E_{\vv{k}n\pm}^{(0)}(\vv{q}) = \pm \sqrt{(\varepsilon_n(\vv{k}) - \mu)^2 + |\mathcal{D}_{n,\vv{k}}(\vv{q})|^2} + \hdots \,,
\end{align}
where we set $\mathcal{D}_{n,\vv{k}} = \bra{\psi_n(\vv{k}-\vv{q})} f(\vv{k}) \ket{\psi_n(\vv{k}+\vv{q})}$. Here, $\ket{\psi_n}$ denotes the $n$th eigenstate of the single-particle Hamiltonian with eigenvalue $\varepsilon_n$. Note that higher order terms are zero for the flat band and are not relevant for the isolated (dispersive) bands. According to the chain rule, we have
\begin{align}
    \frac{\partial^2}{\partial q_i \partial q_j} \klr{-T \ln(1 + \exp\klr{-\frac{E_{\vv{k}n\pm}^{(0)}(\vv{q})}{T}})}\Bigg|_{\vv{q}=0} &=  n_{\mathrm{F}}'(E_{\vv{k}n\pm}^{(0)}(\vv{q})) \frac{\partial E_{\vv{k}n\pm}^{(0)}(\vv{q})}{\partial q_i} \frac{\partial E_{\vv{k}n\pm}^{(0)}(\vv{q})}{\partial q_j} + n_{\mathrm{F}}(E_{\vv{k}n\pm}^{(0)}(\vv{q})) \frac{\partial^2 E_{\vv{k}n\pm}^{(0)}(\vv{q})}{\partial q_i \partial q_j}\klsr{}_{\vv{q}=0} \\
    &\approx  -\frac{4n_{\mathrm{F}}(E_{\vv{k}n\pm}^{(0)})}{E_{\vv{k}n\pm}^{(0)}}|f(\vv{k})|^2 g_{ij}^{(n)}(\vv{k})  \,,
\end{align}
where we used $\partial_{q_i}|\mathcal{D}_{n,\vv{k}}(\vv{q})|^2|_{\vv{q}=0} = 0$ due to the uniform pairing condition. We can neglect the curvature of the other nonflat bands because they are well separated from the flat band such that $E_{\vv{k}n}^{(0)} \gg E_{\vv{k}} \equiv E_{\vv{k}n_0}^{(0)}$ for $n \neq n_0$. By using $n_{\mathrm{F}}(-E) = 1 - n_{\mathrm{F}}(E)$, we can approximate the geometrical superfluid weight by
\begin{align}
    D_{\mathrm{s},ij}^{\mathrm{geom}}(T) \approx \frac{4}{V} \sum_{\vv{k}} \frac{{1 - 2n_{\mathrm{F}}(E_{\vv{k}})}}{E_{\vv{k}}} |f(\vv{k})|^2 g_{ij}(\vv{k}) \,,
    \label{eq:geom-weight-isolated-flat-bands}
\end{align}
where $g_{ij} \equiv g_{ij}^{(n_0)}$ represents the quantum metric of the $n_0$th band~\cite{provost1980riemannian}. This expression is used to calculate the temperature dependence of the geometrical superfluid weight. Analogously, we have
\begin{align}
    \frac{\delta}{\delta f_\alpha^\mu(\vv{q};\vv{k}')}\klr{\frac{\partial}{\partial q_i} \klr{-T \ln(1 + \exp\klr{-\frac{E_{\vv{k}n\pm}^{(0)}(\vv{q})}{T}})}}\klsr{}_{\vv{q}=0} &\approx 
    \frac{{2} n_{\mathrm{F}}(E_{\vv{k}n\pm}^{(0)}) {s_\mu}{f}^{\bar\mu}(\vv{k})}{E_{\vv{k}n\pm}^{(0)}} \sum_{m\neq n} \mathrm{Im}\!{\kle{\braket{\psi_n}{\alpha}\braket{\alpha}{\psi_m} \braket{\psi_m}{\partial_{k_i}\psi_n}}\,,}
\end{align}
where we define $s_{\mathrm{R},\mathrm{I}} = \pm 1$, set $\bar{\mu} = \mathrm{R}$ for $\mu = \mathrm{I}$ and $\bar{\mu} = \mathrm{I}$ for $\mu = \mathrm{R}$, and denote the standard basis of the Hilbert space by $\klg{\ket{\alpha}}$. When neglecting the other separated bands, we obtain the following expression for the functional contribution near zero temperature, 
\begin{align}
    D_{\mathrm{s},ij}^{\mathrm{func}}(T) \approx \frac{4}{V} \sum_{\vv{k},\vv{k}',\alpha,\beta,\mu,\nu} \frac{M_{\alpha\mu,\beta\nu}^{-1}(\vv{k},\vv{k}') {s_\mu s_\nu} f^{\bar{\mu}}(\vv{k}) f^{\bar{\nu}}(\vv{k}') {(2n_{\mathrm{F}}(E_{\vv{k}})-1) (2n_{\mathrm{F}}(E_{\vv{k}'})-1)}}{E_{\vv{k}}E_{\vv{k}'}} h_{ij,\alpha\beta}(\vv{k},\vv{k}') \,,
    \label{eq:func-weight-isolated-flat-bands}
\end{align}
where 
\begin{align}
    h_{ij,\alpha\beta}(\vv{k},\vv{k}') = \frac{1}{2} \sum_{m,m'\neq n_0} \mathrm{Re}\!\kle{O^{(1)}_{n_0n_0mm',\alpha\beta}(\vv{k},\vv{k}') e^{(n_0)}_{i,m}(\vv{k}) \bar{e}^{(n_0)}_{j,m'}(\vv{k}') + O^{(2)}_{n_0n_0mm',\alpha\beta}(\vv{k},\vv{k}') e^{(n)}_{i,m}(\vv{k}) \bar{e}^{(m')}_{j,n_0}(\vv{k}')}
    \label{eq:multi-state-quantum-geometric-term}
\end{align}
represents a multi-state quantum-geometric quantity in Wilczek-Zee representation. Here, $e^{(n)}_{i,m}(\vv{k}) = i\braket{\psi_m(\vv{k})}{\partial_{k_i} \psi_n(\vv{k})}$ is the Wilczek-Zee connection and $O^{(1,2)}_{nn'mm',\alpha\beta}$ are coefficients depending only on the Bloch components. Their explicit expressions are given in Ref.~\cite{buthenhoff2025functional}, while they are not needed for our discussion. Moreover, the Hessian matrix $M$ is obtained as
\begin{align}
    M_{\alpha\mu,\beta\nu}(\vv{k},\vv{k}') = \sum_{\vv{k}'',n,\sigma} \Bigg[ n_{\mathrm{F}}'(E_{\vv{k}''n\sigma}^{(0)}(\vv{q})) \frac{\delta E_{\vv{k}''n\sigma}^{(0)}(\vv{q})}{\delta f^\mu_\alpha(\vv{q};\vv{k})} \frac{\delta E_{\vv{k}''n\sigma}^{(0)}(\vv{q})}{\delta f^\nu_\beta(\vv{q};\vv{k}')} + n_{\mathrm{F}}(E_{\vv{k}''n\sigma}^{(0)}(\vv{q})) \frac{\delta^2 E_{\vv{k}''n\sigma}^{(0)}(\vv{q})}{\delta f^\mu_\alpha(\vv{q};\vv{k}) \delta f^\nu_{\beta}(\vv{q};\vv{k}')}\Bigg]\Bigg|_{\vv{q}=0} + VU_{\mu\nu}^{-1}(\vv{k},\vv{k}') \delta_{\alpha\beta} \,,
    \label{eq:2nd-derivative}
\end{align} 
where $U_{\mu\nu}(\vv{k},\vv{k}') = \mathrm{Re}(U(\vv{k},\vv{k}')) \delta_{\mu\nu} - \mathrm{Im}(U(\vv{k},\vv{k}')) \epsilon_{\mu\nu}$. In particular, by making use of
\begin{align}
    \frac{\delta E_{\vv{k}''n\sigma}^{(0)}(\vv{q})}{\delta f_{\alpha}^\mu(\vv{q};\vv{k})}\klsr{}_{\vv{q}=0} &= \frac{|\braket{\alpha}{\psi_n}|^2f^\mu(\vv{k})}{E_{\vv{k}n\sigma}^{(0)}} \delta_{\vv{k}\vv{k}''}\,, \\
    \frac{\delta^2 E_{\vv{k}''n\sigma}^{(0)}(\vv{q})}{\delta f^\mu_\alpha(\vv{q};\vv{k}) \delta f^\nu_{\beta}(\vv{q};\vv{k}')}\klsr{}_{\vv{q}=0} &= \klr{\frac{|\braket{\alpha}{\psi_n}|^2 \delta_{\alpha\beta}\delta_{\mu\nu}}{E_{\vv{k}n\sigma}^{(0)}} - \frac{|\braket{\alpha}{\psi_n}|^2 |\braket{\beta}{\psi_n}|^2 f^\mu(\vv{k}) f^\nu(\vv{k})}{\big(E_{\vv{k}n\sigma}^{(0)}\big)^3}} \delta_{\vv{k}\vv{k}''} \delta_{\vv{k}'\vv{k}''}
\end{align}
and defining
\begin{align}
    \Pi_{\alpha\mu,\beta\nu}(\vv{k}) =&  2n_{\mathrm{F}}'(E_{\vv{k}}) \frac{|\braket{\alpha}{\psi_n}|^2|\braket{\beta}{\psi_n}|^2 f^\mu(\vv{k})f^\nu(\vv{k})}{\big(E_{\vv{k}}\big)^2} \nonumber \\
    &+ 
    (2n_{\mathrm{F}}(E_{\vv{k}})- 1) \klr{
    \frac{|\braket{\alpha}{\psi_n}|^2 \delta_{\alpha\beta}\delta_{\mu\nu}}{E_{\vv{k}}} - \frac{|\braket{\alpha}{\psi_n}|^2 |\braket{\beta}{\psi_n}|^2 f^\mu(\vv{k}) f^\nu(\vv{k})}{\big(E_{\vv{k}}\big)^3}
    }
    \label{eq:pi-def}
\end{align}
with the Sherman-Morrison-Woodbury formula~\cite{woodbury1950inverting}, we find the inverse of $M$ to correspond to a geometric series as
\begin{align}
    M^{-1}_{\alpha\mu,\beta\nu}(\vv{k},\vv{k'}) =& \frac{1}{V} U_{\mu\nu}(\vv{k},\vv{k}') \delta_{\alpha\beta} \nonumber \\
    &+ \frac{1}{V} \sum_{n=1}^\infty \frac{1}{V^n} \sum_{\substack{\mu_1,\hdots,\mu_n=\mathrm{R},\mathrm{I}\\\nu_1,\hdots,\nu_n=\mathrm{R},\mathrm{I}\\\vv{k}_1,\hdots,\vv{k}_n\\\alpha_1,\hdots,\alpha_{n-1}}} U_{\mu\nu_1}(\vv{k},\vv{k}_1) \Pi_{\alpha\nu_1,\alpha_1\mu_1}(\vv{k}_1) U_{\mu_1\nu_2}(\vv{k}_1,\vv{k}_2) \hdots \Pi_{\alpha_{n-1}\nu_{n},\beta\mu_n}(\vv{k}_n) U_{\mu_n\nu}(\vv{k}_n,\vv{k}') \,.
\end{align}
\twocolumngrid

\bibliography{apssamp}

\end{document}